\documentclass[5p,twocolumn,times,number]{elsarticle}
\usepackage{graphicx}
\usepackage{amsmath}
\usepackage{subfig}
\usepackage{lineno}
\begin{document}
\begin{frontmatter}
  \title{High Rate Resistive Plate Chamber for LHC detector upgrades}
  
  \author[add1]{Y.~Haddad\corref{cor}}
  \ead{haddad@llr.in2p3.fr}
  \author[add2]{I.~Laktineh}
  \author[add2]{G.~Grenier}
  \author[add2]{N.~Lumb}
  \author[add3]{S.~Cauwenbergh}
  
  \cortext[cor]{Corresponding author}
  
  \address[add1]{Laboratoire Leprince-Ringuet (LLR), \'Ecole Polytechnique, 91120 Palaiseau, France}
  \address[add2]{IPNL, Villeurbanne 69622 Lyon, France}
  \address[add3]{Ghent University, Ghent, Belgium}
  
  \begin{abstract}
    The limitation of the detection rate of standard bakelite resistive plate chambers (RPC) used as muon detectors in the LHC
    experiments has prevented the use of such detectors in the high rate regions in both CMS and ATLAS detectors.
    One alternative to these detectors are RPCs made with low resistivity glass plates
    ($10^{10}~{\rm \Omega .cm}$), a beam test at DESY has shown that such detectors can 
    operate at few thousand Hz/cm$^2$ with high efficiency( $> 90\%$).
  \end{abstract}
  

\end{frontmatter}


\section{Introduction}
RPCs are powerful detectors used in many HEP physics experiments. Their
good time resolution and efficiency, in addition to their simplicity and low cost
make them excellent candidates for very large area detectors. The high resistivity of glass
plates helps to prevent discharge damage in these detectors, but this feature represents a
weakness when it comes to their use in high rate environments.

A semi-conductive glass RPC (GRPC) is a solution to overcome this issue. The low resistivity of its
doped glass accelerates the absorption of the avalanche's charge created when a charged
particles crosses the RPC. A recent beam test at DESY in January 2012 with a high rate electron
beam constitutes a validation of this new concept.

The GRPC detector is based on the ionization produced by charged particles in a gas gap. A 
typical gas mixture is $93\%$ TFE($\rm C_2F_4$), $5\%$ $\rm CO_2$ and $2\%$ $\rm SF_6$, contained in a $1.2~\rm mm$ gap
between 2 glass plates. A high voltage between $6.5~\rm kV$ and $8~\rm kV$ was applied on the glass through a
resistive coating, assuring the charge multiplication of initial ionizations in avalanche mode with a typical gain of $10^7$.
\begin{figure}[!h]
  \centering
  \subfloat[]{\label{fig:design-a}\includegraphics[width=0.95\linewidth]{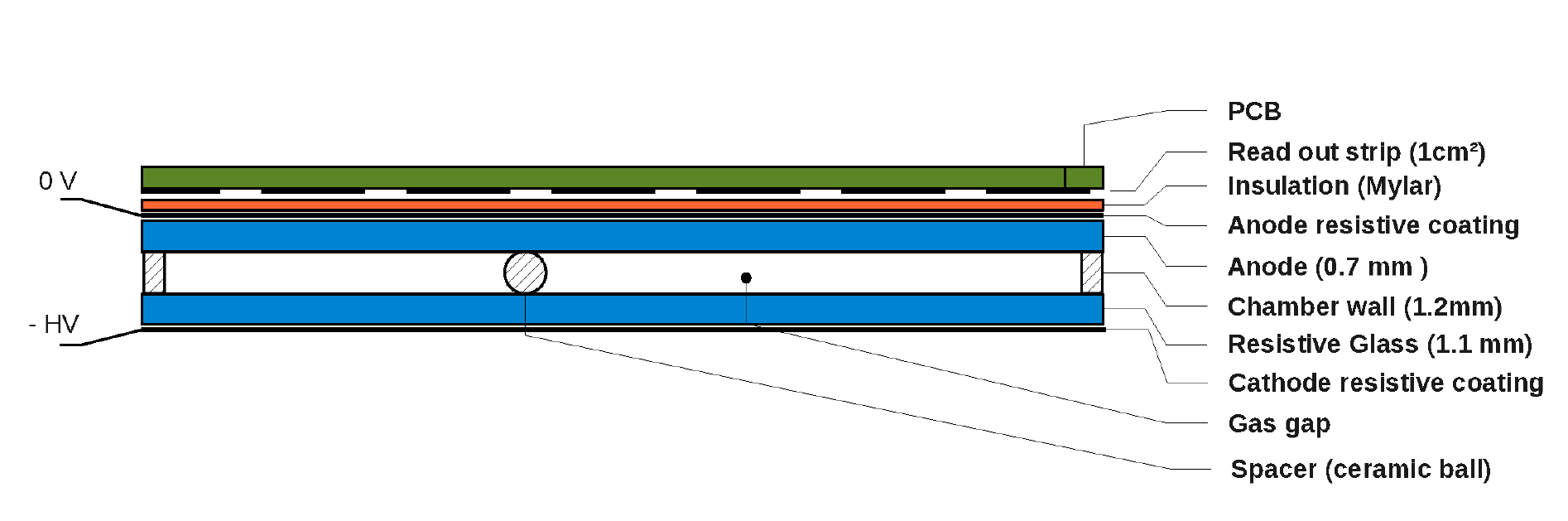}}\\
  \subfloat[]{\label{fig:design-b}\includegraphics[width=0.45\linewidth]{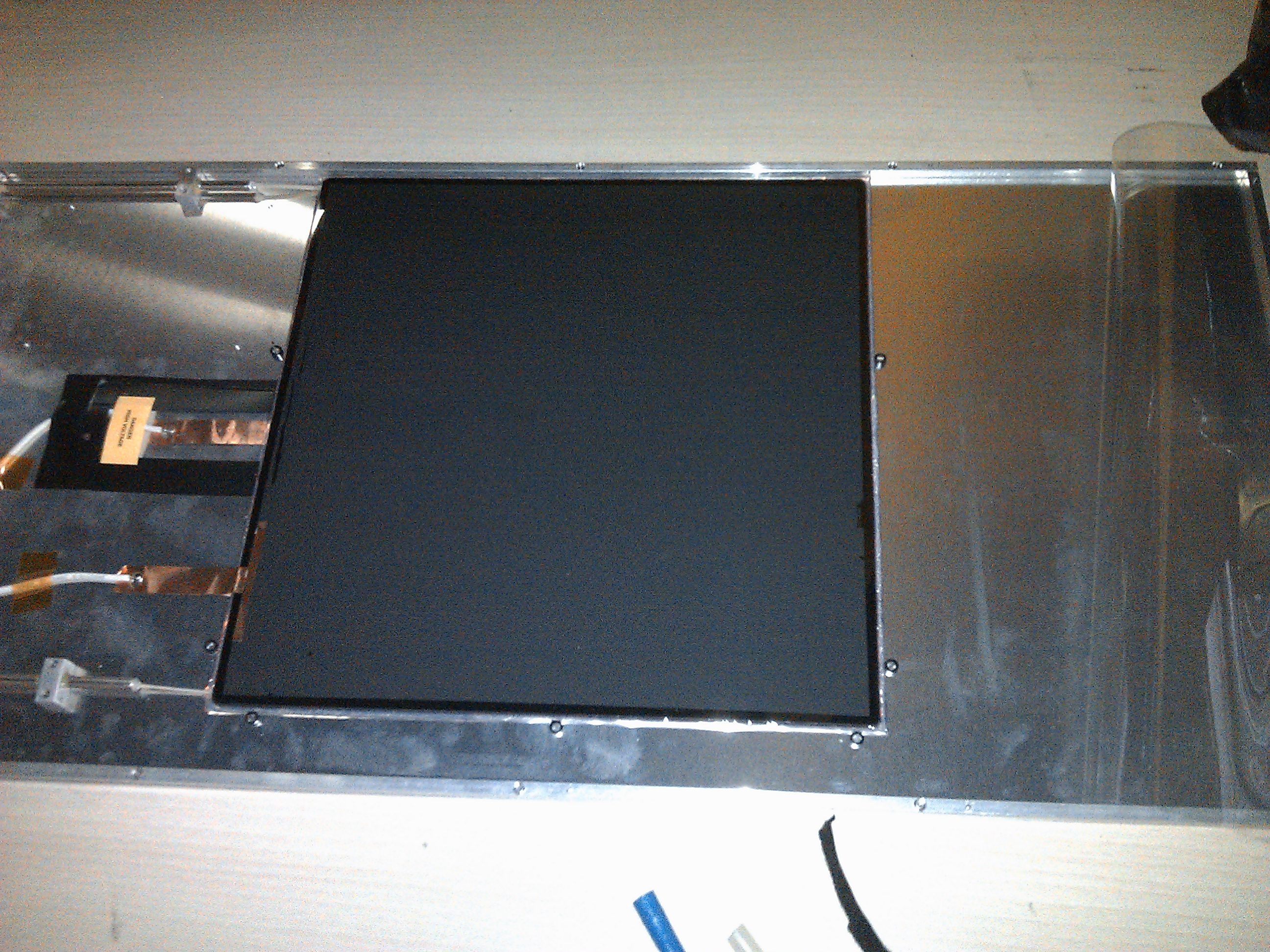}}
  \vspace{0.5cm}
  \subfloat[]{\label{fig:design-c}\includegraphics[width=0.365\linewidth]{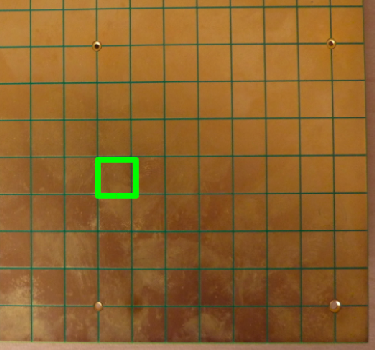}}
  \caption{\footnotesize (a) Schematic drawing of GRPC with electrodes made of silicate glass. (b)
photo of $\rm 30~cm\times 30~cm$ GRPC with semi-conductive glass. (c) readout pad with size of $1 \times 1 \rm cm^2$}
  \label{fig:design.}
\end{figure}
The new aspect of this detector is the low resistivity  of the doped silicate glass (less then $10^{10-11}\rm \Omega.cm$, compared to the $10^{13}\rm \Omega.cm$ typical of float glass), provided by
Tsinguha University following a new process \cite{Wang}.
The glass plate thickness is $1.1~\rm mm$ for the cathode and $0.7~\rm mm$ for the anode. The resistive coating is colloidal graphite of 
$1~M\Omega/\Box$ resistivity. The gas was uniformly distributed in the chamber using the channeling-based system. Ceramic balls with $1.2~\rm mm$ diameter were used 
as spacers. The total GRPC thickness was $3~\rm mm$. The signal was collected by $1 \times 1~\rm cm^2$ copper pads (figure \ref{fig:design-c}) connected to a semi-digital 
readout system with 3 thresholds, identical to the one equipping the GRPC chambers used in the SDHCAL prototype developed within the CALICE 
collaboration \cite{Imad1}\cite{Imad2}.
\section{DESY test beam}
Four $30\times 30~\rm cm^2$ area RPCs were built following the design shown in figure \ref{fig:design-a} and were tested at DESY in January 2012. 
The DESY II synchrotron provides an intense and continuous electron beam with an energy up to $6~\rm GeV$. The particle rate depends on the beam energy,with a 
maximum of $35~\rm kHz$. The beam size is a few $\rm cm^2$. Two scintillator detectors were placed upstream of the detector. Their 
role was to measure the beam rate.  
\begin{figure}[!h]
  \centering
  \label{fig:desy}\includegraphics[width=0.5\linewidth]{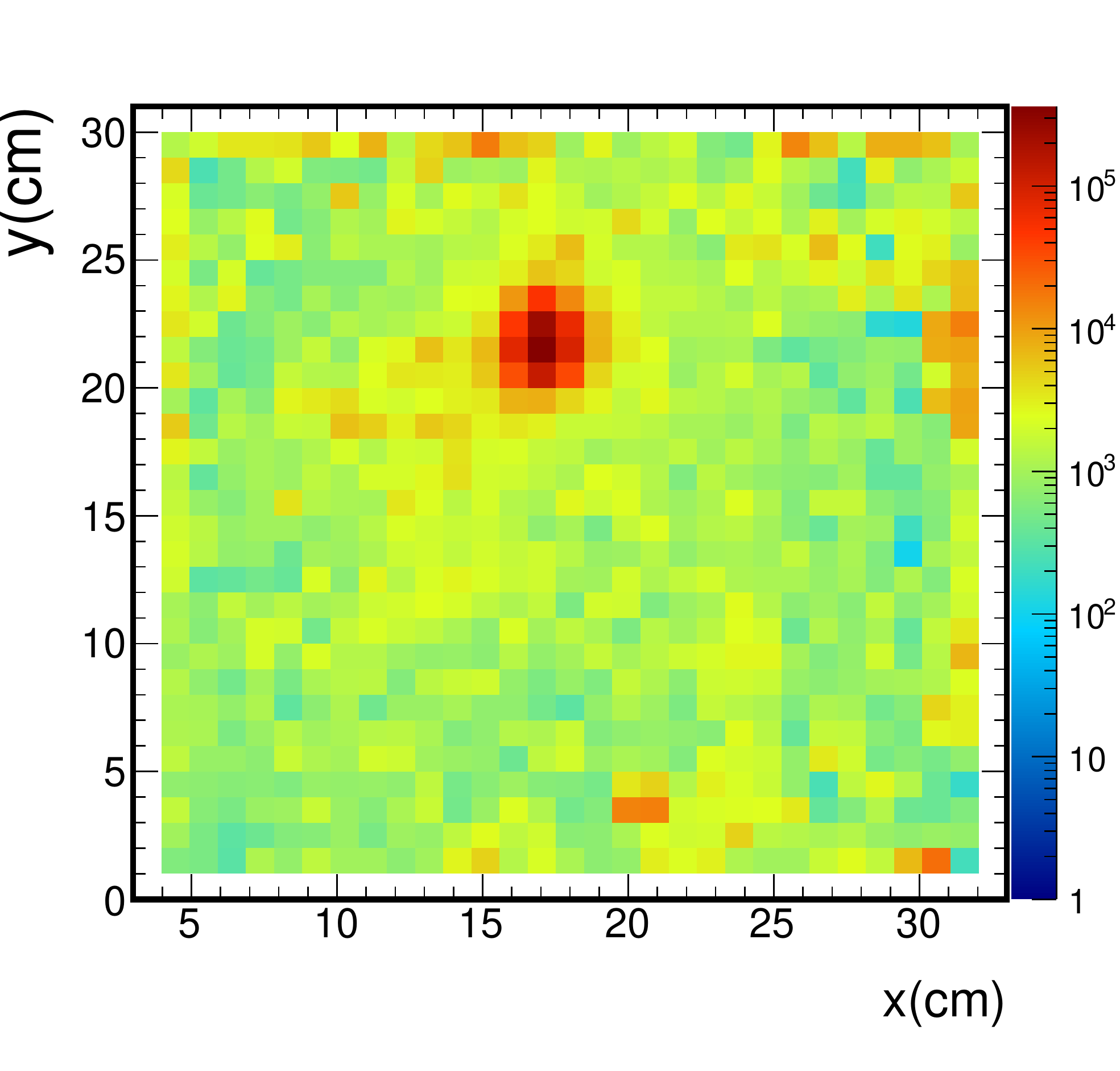}
  \caption{\footnotesize Beam profile in the chambers with $e^{-}$ at $2~\rm GeV$ .}
\end{figure}
One additional GRPC made with standard glass was added to the setup.
\section{Results \& discussion}
\subsection{GRPC performances}
\begin{figure}[!h]
  \centering
  \subfloat[]{\label{fig:desy-a}\includegraphics[width=0.52\linewidth]{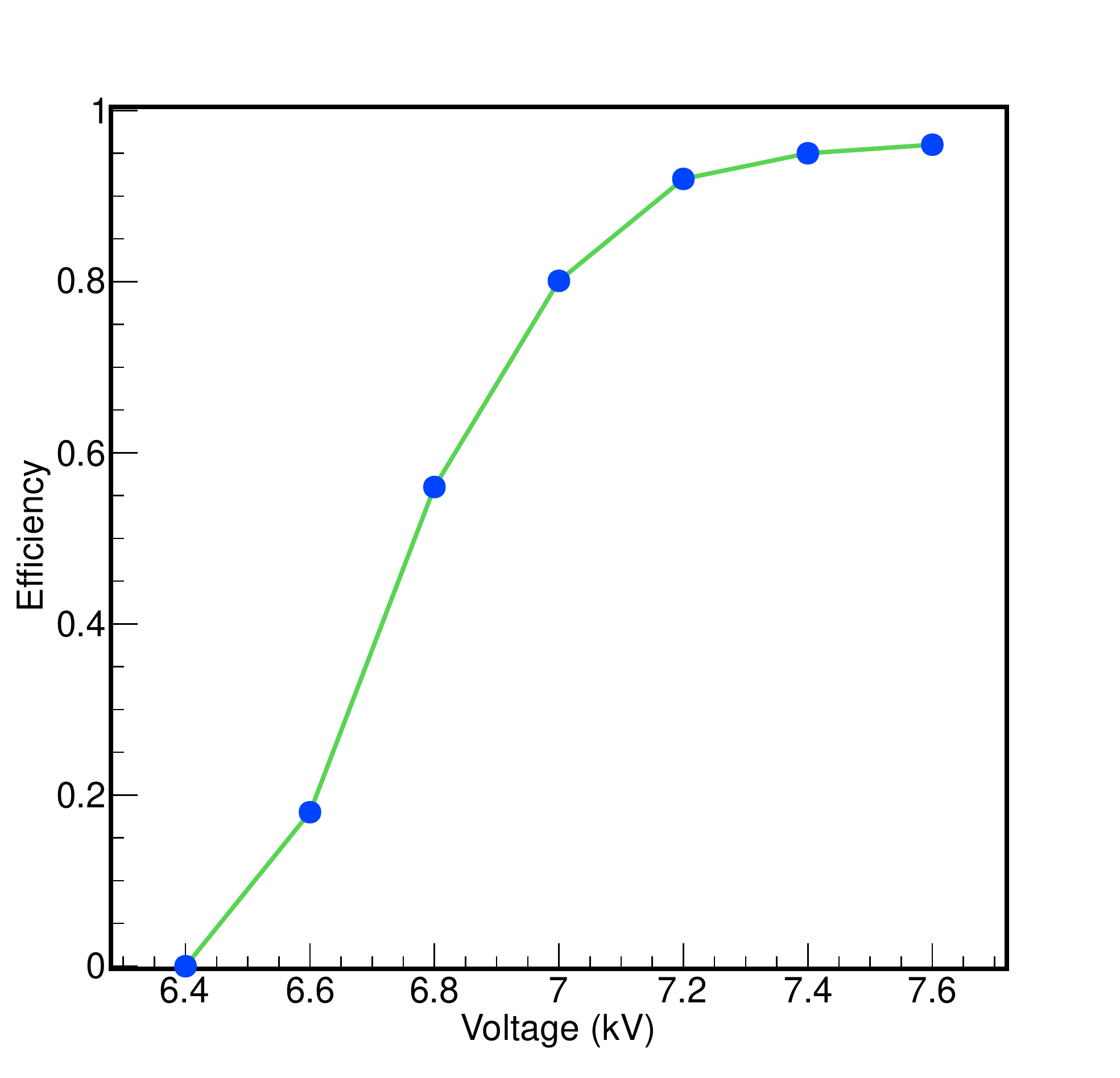}}
  \subfloat[]{\label{fig:desy-b}\includegraphics[width=0.52\linewidth]{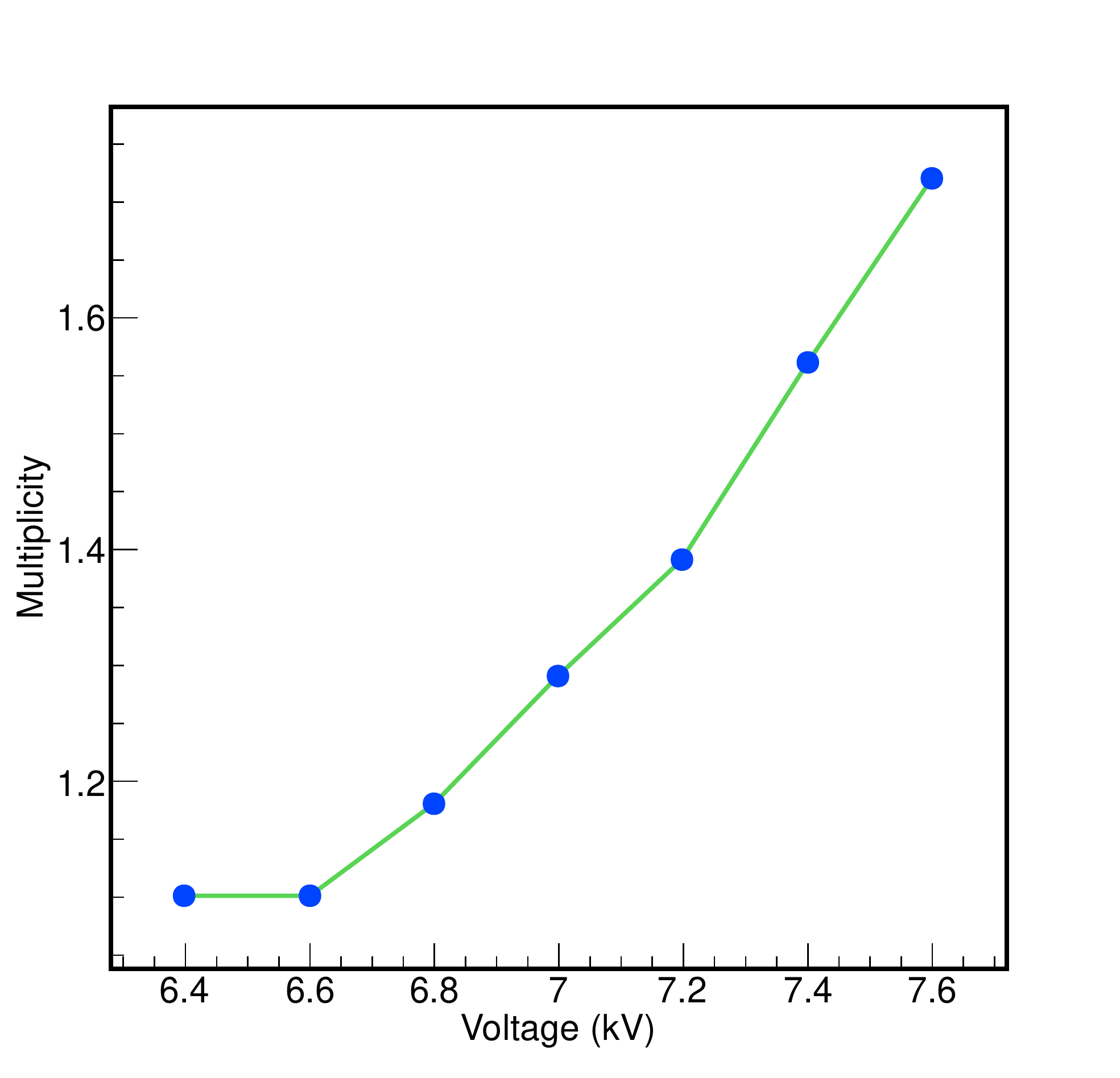}}
  \caption{\footnotesize (a) Efficiency vs high voltage scan. (b) Multiplicity vs high voltage scan}
\end{figure}

The local efficiency and multiplicity were measured by using 3 chambers to reconstruct particle tracks and determining the expected 
hit position in the 4th. The multiplicity $\rm \mu$ is defined as the number of fired pads within $3~cm$ of the expected position. 
The efficiency $\rm \epsilon$ is the fraction of tracks with $\mu \geq 1$. The efficiency (\ref{fig:desy-a})  and multiplicity
(\ref{fig:desy-b}) were measured as function of the polarization high voltage. The same threshold was used for all voltages. The threshold value is fixed  
at $50~\rm fC$ and $7.2~ \rm kV$ was chosen as the working point, giving $(\mu,\epsilon) = (1.4~,~95\%)$. 

\subsection{Running in a high rate beam}
\begin{figure}[!h]
  \centering
  \includegraphics[width=0.75\linewidth]{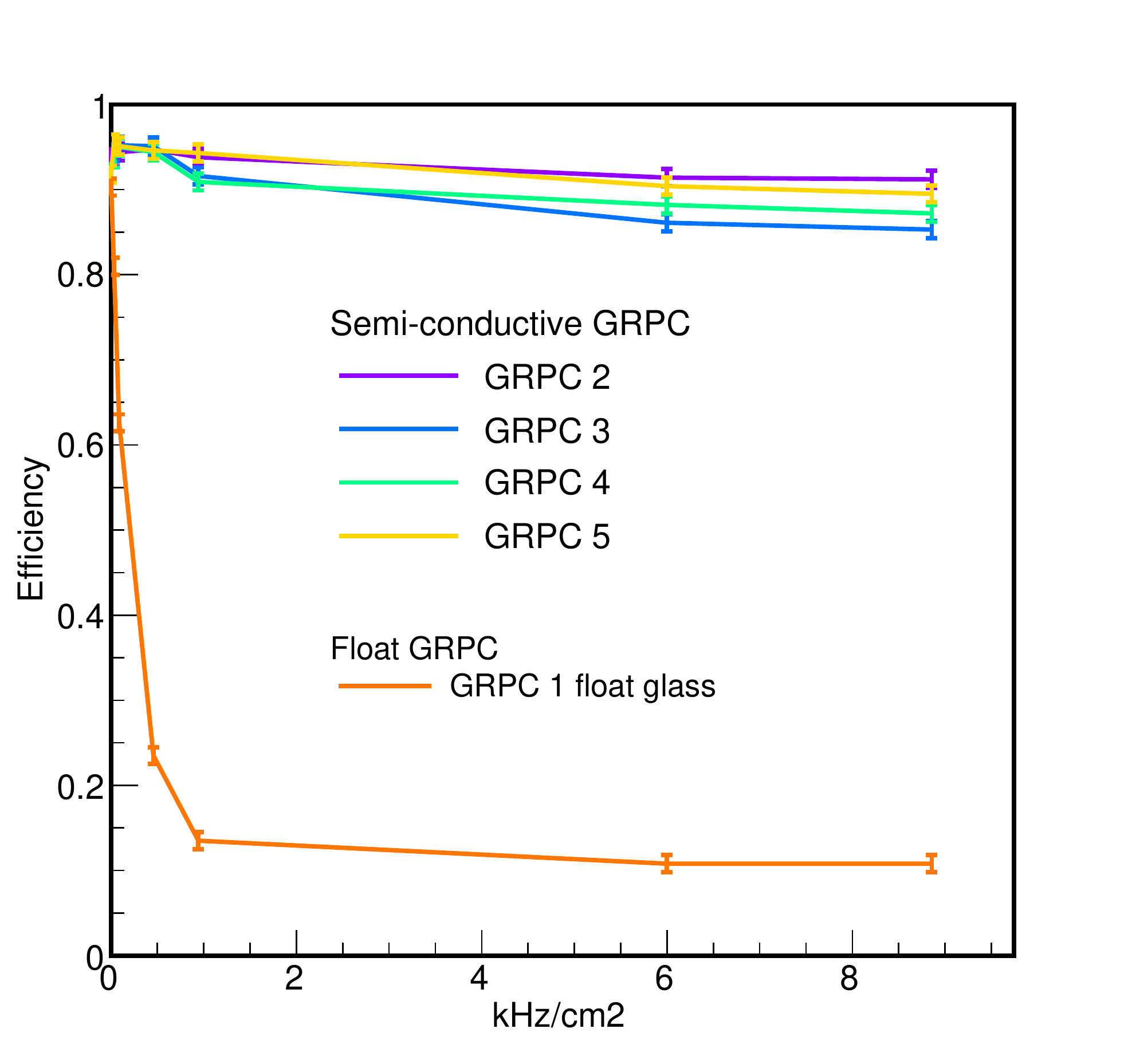}
  \caption{\footnotesize Efficiency vs rate for different RPC. The orange line correspond to the
 GRPC with float glass. The semi-conductive chambers are represented with different colors.}
  \label{fig:EFFRate}
\end{figure}

The scintillator detectors were used to determine the total particle flux, which was then divided by the beam RMS 
area ($\approx 4~\rm cm^2$) to obtain the rate by unit area. The measured ($\mu$, $\epsilon$) 
for different beam rates are plotted in figure \ref{fig:EFFRate}.
The chamber with standard float glass (GRPC 1) becomes inefficient at rate exceeding one hundred $\rm Hz/cm^2$ (above $1~\rm kHz$ the efficiency is below $10\%$) while the semi-conductive chambers (GRPC 2-5) maintain a high efficiency $\geq 90\%$ until at least $9~\rm kHz/cm^2$. 
\section{Conclusion}
Semi conductive glass RPCs were tested at DESY in a high rate electron
beam, producing very encouraging results; it has been shown that the
main weakness of standard RPCs, namely the drop of efficiency at high
rate, is clearly overcome, with efficiencies remaining at around
$90\%$ at rate of $9~\rm kHz/cm^2$. 
This feature, combined with GRPC capability to provide precise time
measurement, makes them an excellent candidate for future LHC muon
detector upgrades. 
Additional studies on their aging under high rate conditions are
underway. A multi-gap version is also under investigation.

\end{document}